\documentclass[twocolumn,showpacs,preprintnumbers,amsmath,amssymb]{revtex4}
\usepackage{amsmath}
\usepackage{graphicx}% Include figure files
\usepackage{dcolumn}% Align table columns on decimal point
\usepackage{bm}% bold math

\begin{document}
\title{Undamped Bloch Oscillations in the $U\rightarrow \infty$ one-dimensional Hubbard model}
\author{Yong Zheng}
\email{zhengyongsc@sina.com}
\address{School of Physics and Electronics, Qiannan Normal University for Nationalities,
Duyun 558000, China}
\date{Received \today }

\begin{abstract}
The $U\rightarrow +\infty$ one-dimensional
Hubbard model in an electric field has be exactly solved, with an emphasis on the charge current. It is found that undamped Bloch
oscillations extensively exist in the system. Such conclusion has also been discussed for  more general cases and we find that it is closely related to the temporal periodicity of the  model Hamiltonian in electric field, rather than to the 
 integrability of the model. As a comparison, we have also studied a model
of electrons with $\delta$-function interactions in continuous space, which is closely
related to the Hubbard model, but is non-integrable; and we find that the charge current strangely  shows a dissipationless behaior which is comparable with the undamped Bloch oscillations.
\end{abstract}


\maketitle

\section{Introduction}

Though one-dimensional (1D) Hubbard model can be exactly solved with Bethe ansatz
\cite{Bt1,Bt2}, the discussion of its response to electric field does not benefit much from this.
In the simplest treatment, the electric field can be introduced to the model via  a time-dependent Peierls phase factor \cite{PS,PS2}, even though, an exact addressing of the response behavior of the system generally is still impossible.

To the level of
linear response, many studies have concluded that 1D Hubbard model shows ideal conductance  \cite{In1,In2,In3,In4}; and such
behavior is closely related to the integrability of the model, which
always results in a nonzero Drude weight in cases away from half filling. 

While to the level of nonlinear response, less is known about this model. Generally, it is believed that the integrability of a model still plays an important role  \cite{cc,cc1} in a nonlinear response: for integrable models, the charge current can resemble
the Bloch oscillations as in the noninteracting
 case; however, the appearance of electric field may  break integrability of the model itself,  resulting in 
 a damped current oscillation. 
 
 Actually, as for Hubbard model, Eckstein and Werner \cite{1}, using dynamical mean-field theory,  have shown numerically that,  
 different values of the on-site Coulomb repulsion $U$ of electrons can lead to different damped-oscillating behaviors of charge current. And the
 oscillating feature would even disappear if $U$ exceeds some critical
 value.  It seems that the damping of Bloch oscillations is
 inevitable in the presence of  on-site $U$. Such picture has been further supported by Mandt \cite{2} via
 a variational solution of the Boltzmann equation for the Hubbard
 model. 
 
Nevertheless, since the exact solution of the Hubbard model in electric field is  lack even in the 1D case, 
 it is still an open problem that how the model responses an electric field nonlinearly. However, we find that 1D Hubbard model in the 
 $U\rightarrow +\infty$  limit provides a good platform  to address this question. We will show that in such case, using a method of unitary transformation, the model is still solvable when the electric field appearing.
 
We will show that the behavior of charge current indeed can resemble
  Bloch oscillations, but such behavior is closely related to the temporal periodicity of the  Hamiltonian in electric field, rather than to the integrability of the model. And this conclusion can be extended to more general models and cases. 
 
 Especially, to see more about 
 the effect of integrability and 
  correlation interactions on the response of a
 system to electric field,  we will show in the appendix a non-integrable model of electrons with $\delta$-function interactions, of which, however, the conductivity can be exactly determined.

\section{Model and transformation}
We consider an $N$-electron system on a periodic ring with $L$ sites, and we can treat the constant electric field  as the temporal derivative of a time-dependent vector
potential along the ring, $E=-1/c A_t (t)$ or ${A} (t)=-cEt$.
The model Hamiltonian then can be written as
\begin{equation} \label{H1}
H=-t_{0}\sum_{j,\sigma}[e^{i\frac{e}{\hbar
c}A(t)}C^{\dag}_{j\sigma}C^{
}_{j+1\sigma}+\text{H.c.}]+U\sum_{j}n_{j\uparrow}n_{j\downarrow},
\end{equation}
where $C^{\dag}_{j\sigma}$ ($C^{}_{j\sigma}$) is  the creation
(annihilation) operator of electron at site $j$ and spin
$\sigma=\uparrow$ or $\downarrow$. And $n_{j\sigma}$ stands for the
particle number operator as usual.  We use the periodic boundary
condition $C^{\dag}_{j+L\sigma}=C^{\dag}_{j\sigma}$ and
$C^{}_{j+L\sigma}=C^{}_{j\sigma}$. For simplicity,  we set the
lattice constant $ a=1 $.

In the limit $U\rightarrow +\infty$, which we are interested in, the double occupation of a site by $\uparrow$ and $\downarrow$ electrons is prohibited,
and the Hamiltonian is formally reduced to
\begin{equation} \label{H2}
H=-t_{0}\sum_{j,\sigma}[e^{i\frac{e}{\hbar
c}A(t)}\hat{C}^{\dag}_{j\sigma}\hat{C}^{ }_{j+1\sigma} +
\text{H.c.}],
\end{equation}
where
$\hat{C}^{\dag}_{j\sigma}=(1-n_{j\bar{\sigma}})C^{\dag}_{j\sigma}$
and $\hat{C}^{}_{j\sigma}=(1-n_{j\bar{\sigma}})C^{}_{j\sigma}$ as usual.
Then,  we can concentrate our  discussion in the subspace without
double occupations. And the basis states involved can be written in the form   $|\mathbf{j},\boldsymbol{\sigma}\rangle
=C^{\dag}_{j_1\sigma_{1}}C^{\dag}_{j_2\sigma_{2}}\cdots
C^{\dag}_{j_N\sigma_{N}}|0\rangle\equiv
|j_1\sigma_{1},j_2\sigma_{2},\cdots j_N\sigma_{N} \rangle$,
  in which $N$ electrons with spins $\sigma_{1}$,
$\sigma_{2}$, $\cdots$, $\sigma_{N}$ occupy sites
$j_{1}$, $j_{2}$, $\cdots$, $j_{N}$ respectively, where $1\leq
j_{1}<j_{2}<\cdots<j_{N}\leq L$, to  guarantee that no double occupation occurs. And the charge current operator \cite{PS2,co}  can  be
written as
\begin{equation} \label{co}
\widehat{j}=-c\frac{\partial H}{\partial A}.
\end{equation} 

Without electric field, the wave function of 1D
$U\rightarrow +\infty$  Hubbard model has a spin-charge separated
form, as is well-known \cite{Bt2}.  However, when electric field appearing, our problem becomes a time-dependent
 one, and still requires a careful discussing. 

As has be noted previously \cite{89},  all the terms in
Eq.~(\ref{H2}), except the ones with
$\hat{C}^{\dag}_{L\sigma}\hat{C}^{ }_{1\sigma}$ or
$\hat{C}^{\dag}_{1\sigma}\hat{C}^{ }_{L\sigma}$,  when acting on a
 state, can only change the sites of the $N$ electrons, without any alteration 
 in the spin sequence;  that is, they can only
change a state $|j_1\sigma_{1},j_2\sigma_{2},\cdots j_N\sigma_{N} \rangle$
to something like $|j'_1\sigma_{1},j'_2\sigma_{2},\cdots
j'_N\sigma_{N} \rangle$. Actually, without electric filed, basing on this point, the
$U\rightarrow +\infty$ Hubbard model can be  mapped to a set
of Hamiltonians  of spinless
fermions, and can be exactly solved \cite{89,90}. Here, we also base
our discussion on this point, but use a method of unitary
transformation.

Firstly, we introduce a new kind of   basis states, in which the spin and charge are completely independent,
\begin{equation}\label{T1}
|\mathbf{j}\rangle \bigotimes|\boldsymbol{\sigma} \rangle =
|j_1,j_2,\cdots j_N \rangle \bigotimes| \sigma_{1} \sigma_{2}\cdots
\sigma_{N} \rangle,
\end{equation}
where $|\mathbf{j}\rangle = |j_1,j_2,\cdots j_N \rangle$ denotes a
state that $N$ spinless but charged fermions occupy sites $j_{1}$,
$j_{2}$, $\cdots$, $j_{N}$, and it is required that $1\leq
j_{1}<j_{2}<\cdots<j_{N}\leq L$, as before; by introducing the
$j$-site creation (annihilation) operator $f^{\dag}_{j}$
($f^{}_{j}$) of  these fermions,  such state can be expressed as
$|\mathbf{j}\rangle = f^{\dag}_{j_1}f^{\dag}_{j_2}\cdots
f^{\dag}_{j_N}|0\rangle$. And $|\boldsymbol{\sigma} \rangle = |
\sigma_{1} \sigma_{2}\cdots \sigma_{N} \rangle$ is the configuration  of
$N$ sequenced spins. Obviously, the direct product states defined by
Eq.~(\ref{T1}) have a one-to-one correspondence with
$|\mathbf{j},\boldsymbol{\sigma}\rangle$. We can introduce a
transformation
\begin{equation}\label{Tr}
T=\sum_{\mathbf{j},\boldsymbol{\sigma}} |\mathbf{j}\rangle
\bigotimes|\boldsymbol{\sigma} \rangle \langle
\mathbf{j},\boldsymbol{\sigma}|,
\end{equation}
of which the unitary characteristic $T^{\dag}T=\mathbf{1}$ can be easily verified. Then,
we have $T|\mathbf{j},\boldsymbol{\sigma} \rangle=|\mathbf{j}\rangle
\bigotimes|\boldsymbol{\sigma} \rangle $. Under such transformation,
the form of all the terms in Eq.~(\ref{H2}), except the ones with
$\hat{C}^{\dag}_{L\sigma}\hat{C}^{ }_{1\sigma}$ and
$\hat{C}^{\dag}_{1\sigma}\hat{C}^{ }_{L\sigma}$, can be easily
obtained. For $j\neq N$,
\begin{multline*}
 \sum_{\sigma}\hat{C}^{\dag}_{j\sigma}\hat{C}^{ }_{j+1\sigma}\rightarrow
T\sum_{\sigma}\hat{C}^{\dag}_{j\sigma}\hat{C}^{ }_{j+1\sigma}T^{\dag} \\
 =\sideset{}{'}\sum_{\mathbf{j}_i}\widetilde{|\mathbf{j}_i\rangle}\langle \mathbf{j}_i|
\bigotimes\sum_{\boldsymbol{\sigma}}|\boldsymbol{\sigma}
\rangle\langle\boldsymbol{\sigma}|\equiv f^{\dag}_{j}f^{}_{j+1}I,
\end{multline*}
where $|\mathbf{j}_i\rangle$ is a state with site $j$ empty and site
$j+1$ occupied by a spinless fermion. And
$\widetilde{|\mathbf{j}_i\rangle}$ is same to
$|\mathbf{j}_i\rangle$, except with site $j+1$ empty and site $j$
occupied. ``$\Sigma'$'' represents a sum over these states, covering
all the cases in which a spinless fermion hops from site $j+1$ to
site $j$.  With the creation and annihilation operators of spinless
fermions, such sum can be rewritten as $f^{\dag}_{j}f^{}_{j+1}$
\cite{AQ}. And the spin part, due to the unchanged sequence of
spins, is equivalent to a unit operator of spin configurations,
which we have denoted by $I$.

While for  $\hat{C}^{\dag}_{L\sigma}\hat{C}^{ }_{1\sigma}$ or
$\hat{C}^{\dag}_{1\sigma}\hat{C}^{ }_{L\sigma}$, since when acting
on a basis,  they will give a  cyclic permutation of the spin
configuration to the left or right respectively,  we have
\begin{multline*}  %\label{}
\sum_{\sigma}\hat{C}^{\dag}_{L\sigma}\hat{C}^{ }_{1\sigma}\rightarrow
T\sum_{\sigma}\hat{C}^{\dag}_{L\sigma}\hat{C}^{ }_{1\sigma}T^{\dag} \\
=f^{\dag}_{L}f^{}_{1}\sum_{\boldsymbol{\sigma}}\widetilde{|\boldsymbol{\sigma} \rangle}\langle\boldsymbol{\sigma}|\equiv f^{\dag}_{L}f^{}_{1}P,
\end{multline*}
and
\[  %\label{}
\sum_{\sigma}\hat{C}^{\dag}_{1\sigma}\hat{C}^{
}_{L\sigma}\rightarrow
T\sum_{\sigma}\hat{C}^{\dag}_{1\sigma}\hat{C}^{ }_{L\sigma}T^{\dag}=
f^{\dag}_{1}f^{}_{L}P^{-1},
\]
where $\widetilde{|\boldsymbol{\sigma} \rangle}$ represents a spin configuration obtained via  a  left cyclic permutation of $|\boldsymbol{\sigma} \rangle$, that is, if we take $|\boldsymbol{\sigma}\rangle=|\sigma_{1}\sigma_{2}\cdots\sigma_{N}\rangle$, then $\widetilde{|\boldsymbol{\sigma} \rangle}=|\sigma_{2}\cdots\sigma_{N}\sigma_{1}\rangle$. And the sum over all these configurations can be  replaced by the left or right cyclic permutation operator, which we have denoted by $P$ and $P^{-1}$  respectively.

Then the final form of the Hamiltonian after the unitary transformation (\ref{Tr}) is
\begin{multline}\label{hh4}
\widetilde{H}=-t_{0}\sum_{j=1}^{L-1}\left[  e^{i\theta(t)}f^{\dag}_{j}f^{}_{j+1}I+\text{H.c.} \right]  \\
-t_{0}e^{i\theta(t)}f^{\dag}_{L}f^{}_{1}P-t_{0}e^{-i\theta(t)}f^{\dag}_{1}f^{}_{L}P^{-1},
\end{multline}
where $\theta(t)=eA(t)/(\hbar c)=-eEt/\hbar$. 
We will find that our discussion can be
substantially reduced with such new form of Hamiltonian.

And noting that $T$ is independent of $t$ or $A(t)$,  the charge
current operator under this transformation becomes
\begin{equation} \label{cc}
\widetilde{j}=-cT\frac{\partial H}{\partial A}T^{\dag}=-c\frac{\partial
\widetilde{H}}{\partial A}.
\end{equation}

\section{Solutions}

In Eq.~(\ref{hh4}), the  spin-part operators $ P $, $ P^{-1} $ and $
I $ do not cause any trouble for our discussion, since they commute with
each other and the eigenstate can be obtained easily.
 For any spin configuration $|\boldsymbol{\sigma}_{s_1}\rangle=|\sigma_{1}\sigma_{2}\cdots\sigma_{N}\rangle$, we introduce 
$|\boldsymbol{\sigma}_{s_2}\rangle=P|\boldsymbol{\sigma}_{s_1}\rangle$,
$|\boldsymbol{\sigma}_{s_3}\rangle=P|\boldsymbol{\sigma}_{s_2}\rangle=P^2|\boldsymbol{\sigma}_{s_1}\rangle$,
$\cdots$, till some  integer $ m_{s} \leq N$, for which the resulted
state repeats $ |\boldsymbol{\sigma}_{s_1}\rangle $ for the first
time, namely,
$|\boldsymbol{\sigma}_{s_{m_{s}+1}}\rangle=P^{m_{s}}|\boldsymbol{\sigma}_{s_1}\rangle=|\boldsymbol{\sigma}_{s_1}\rangle$.
Obviously, $m_{s}$ is directly related to the detailed form of
$|\boldsymbol{\sigma}_{s_1}\rangle$. These configurations form a
subset as
\begin{equation}\label{s}
s=\{|\boldsymbol{\sigma}_{s_1}\rangle,|\boldsymbol{\sigma}_{s_2}\rangle,\cdots,|\boldsymbol{\sigma}_{s_{m_{s}}}\rangle\},
\end{equation}
 from which $m_{s}$ eigenstates of  $P$ or $P^{-1}$ can be obtained as follows,
\begin{multline}\label{ks}
|\chi_{k_s}\rangle=1/\sqrt{m_s}[ \exp(ik_s)|\boldsymbol{\sigma}_{s_1}\rangle+\exp(i2k_s)|\boldsymbol{\sigma}_{s_2}\rangle \\
 +\cdots+\exp(im_sk_s)|\boldsymbol{\sigma}_{s_{m_s}}\rangle ],
\end{multline}
where $ k_s=2\pi m/m_s $, and $ m=0,1,2,\cdots,m_s-1 $. We have $P|\chi_{k_s}\rangle=\exp(ik_s)|\chi_{k_s}\rangle$, $P^{-1}|\chi_{k_s}\rangle=\exp(-ik_s)|\chi_{k_s}\rangle$.
Furthermore, we obtain the representation  $P=\sum_{s}\sum_{k_s}\exp(ik_s)|\chi_{k_s}\rangle\langle\chi_{k_s}|$, $P^{-1}=\sum_{s}\sum_{k_s}\exp(-ik_s)|\chi_{k_s}\rangle\langle\chi_{k_s}|$ and $I=\sum_{s}\sum_{k_s}|\chi_{k_s}\rangle\langle\chi_{k_s}|$,
where ``$ \sum_{s} $'' is a sum over all the possible subsets. Then, the Hamiltonian (\ref{hh4}) is reduced to
\begin{multline}\label{hh}
\widetilde{H}=\sum_{s}\sum_{k_s}\bigg[ -t_{0}e^{i\theta (t)}\bigg(\sum_{j=1}^{L-1} f^{\dag}_{j}f^{}_{j+1}+f^{\dag}_{L}f^{}_{1}e^{ik_s}\bigg) + \text{H.c.}\bigg] \\
\bigotimes{}(|\chi_{k_s}\rangle\langle\chi_{k_s}|),
\end{multline}
which is  already in a diagonal form now as far as the spin part is
concerned. What we need to do next is just to diagonalize the
spinless-fermion part inside  ``[ ]'', which we can denote by $ h_s $,  associating with each
$|\chi_{k_s}\rangle$. Introducing $
h_{0}=-t_{0}e^{i\theta(t)}
(\sum_{j=1}^{L-1}f^{\dag}_{j}f^{}_{j+1}+f^{\dag}_{L}f^{}_{1}e^{ik_s}
) $ to write $ h_s=h_{0}+h_{0}^{\dagger} $, and noting $ [h_{0},
h_{0}^{\dagger} ] =0 $,  we find that both $ h_{0} $ and $
h_{0}^{\dagger} $, and hence $ h_s $, can be written in a diagonalized
form simultaneously. To do this, we assume a form of  instantaneous
eigenstates for $h_{0}$ as   $ | \varphi \rangle= \sum_{j=1}^{L}
a_{j}f^{\dag}_{j}| 0 \rangle $. Letting  $h_{0} | \varphi \rangle=
a| \varphi \rangle$ yields
\begin{equation}\label{xs}
-t_{0}e^{i\theta(t)} a_{j+1}=aa_{j}, \text{ $j\neq
N$},
\end{equation}
 and
\begin{equation}\label{xs1}
-t_{0}e^{i\theta(t)}\exp(ik_s) a_{1}=aa_{N}.
\end{equation}

From Eq.~(\ref{xs}),
\[ a_{N}=\frac{a}{-t_{0}e^{i\theta(t)}} a_{N-1}=\cdots=(\frac{a}{-t_{0}e^{i\theta(t)}})^{N-1} a_{1},\]
combining which   with Eq.~(\ref{xs1}) yields $a=a_{q}\equiv
-t_{0}e^{i\theta(t)}\exp(iq)$, where
\begin{equation}\label{q}
q=k_s/L+2n\pi/L,\
n=0,1,2,\cdots,L-1.
\end{equation}

For each $q$, repeatedly using (\ref{xs}) to determine the
coefficient $a_j$, we finally obtain an eigenstate which can be
normalized  as $ | \varphi_{q} \rangle=
\frac{1}{\sqrt{L}}\sum_{j=1}^L \exp(iqr_j)| j \rangle =f^{\dag}_{q}|
0 \rangle$, where
\begin{equation}\label{fq}
f^{\dag}_{q}=\frac{1}{\sqrt{L}}\sum_{j=1}^L \exp(iqr_j)f^{\dag}_{j}.
\end{equation}

It can be verified that $h_{0}^{\dagger} | \varphi_{q}
\rangle=a_{q}^{\ast}| \varphi_{q} \rangle$, and then $ h|
\varphi_{q} \rangle= (a_{q}+a_{q}^{\ast})| \varphi_{q}
\rangle=\varepsilon_{q}(t)| \varphi_{q} \rangle$, with
\begin{equation}\label{eq}
\epsilon_{q}(t)=-2t_{0}\cos \left[ q+\theta(t)\right] .
\end{equation}

Further, we can write $h_s$ as $h_s=\sum_{q}\epsilon_{q}(t)| \varphi_{q}
\rangle \langle \varphi_{q}|$ or in an operator form
\begin{equation}\label{hhd}
 h_s=\sum_{q}\epsilon_{q}(t)f^{\dag}_{q}f^{}_{q}.
\end{equation}

And using Eq.~(\ref{fq}), it can be verified  that  $f^{\dag}_{q}$
or $f^{}_{q}$ is still a Fermi operator. The eigenstate of $h$ for
$N$ spinless fermions at each time $t$,  can be written as
\begin{equation}\label{qst}
|\mathbf{q}\rangle \equiv f^{\dag}_{q_1}f^{\dag}_{q_2}\cdots
f^{\dag}_{q_N}|0\rangle,
\end{equation}
 where
 we have used an $ N $-dimensional vector $\mathbf{q}=(q_1,q_2,\cdots,q_N)$ to simplify the formulation. Due to the anticommuting property of $f^{\dag}_{q}$'s, $q_1\neq
q_2\neq \cdots\neq q_N$. Then the $N$-electron eigenstate of
$\widetilde{H}$ at time $t$ can be written as
\begin{equation}\label{st}
|\Psi_{\mathbf{q}s}\rangle=|\mathbf{q}\rangle\bigotimes{}|\chi_{k_s}\rangle,
\end{equation}
with a instantaneous eigenvalue
\begin{equation}\label{E}
E_{\mathbf{q}}(t)=\epsilon_{q_1}(t)+\epsilon_{q_2}(t)+\cdots+\epsilon_{q_N}(t).
\end{equation}

 The most important thing is that, though $\widetilde{H}$ and the
instantaneous eigenvalue $E_{\mathbf{q}}(t)$  both depend on time,
the eigenstate $|\Psi_{\mathbf{q}s}\rangle$ is independent of time.
This permits us an exact study of the time evolving behavior of the
system.  Since any initial state of the system can be represented by
the instantaneous eigenstates at the initial time, we only focus
on the case in which the initial state is an  instantaneous
eigenstate of the system. Namely, we assume that  at $t=0$, the
state of the system, $|\Psi(0)\rangle=|\Psi_{\mathbf{q}s}\rangle$.
Since $|\Psi_{\mathbf{q}s}\rangle$ is still an instantaneous
eigenstate of $\widetilde{H}$ at any time latter, the evolving state
$|\Psi(t)\rangle$ differs from $|\Psi(0)\rangle$ only by a
time-dependent factor and can be determined via
\begin{equation*}\label{pte}
i\hbar\frac{\partial }{\partial
t}|\Psi(t)\rangle=\widetilde{H}|\Psi(t)\rangle=
E_{\mathbf{q}}(t)|\Psi (t)\rangle,
\end{equation*}
following from which,
\begin{equation}\label{pt}
|\Psi(t)\rangle= \exp \Big[\frac{-i}{\hbar}\int_{0}^{t}
E_{\mathbf{q}}(t) dt\Big ]|\Psi_{\mathbf{q}s}\rangle.
\end{equation}

With these time evolution states, we can make a detailed discussion about the time dependent properties of the system. 

Obviously, we can also  apply our solving  strategy  to other similar models in the 1D case, such as the strong-coupling $ t $-$ V $ model \cite{tv} and SU($ N $) model of impenetrable fermions \cite{SU}, to study the time evolution of states in electric field.

\section{Results and discussions}

The charge current, which we are most interested in, is 
\begin{multline}\label{j}
j(t)=\langle\Psi(t)| \widetilde{j}|\Psi(t)\rangle=-c\langle\Psi(t)| \frac{\partial \widetilde{H}}{\partial A}|\Psi(t)\rangle  \\
=-c\frac{\partial }{\partial A} \langle\Psi(t)| \widetilde{H}|\Psi(t)\rangle=-c\frac{\partial E_{\mathbf{q}}(t)}{\partial A},
\end{multline}
where we have used the fact that $\frac{\partial }{\partial A}
\langle\Psi(t)| \Psi(t)\rangle=0$.

Using Eq.~(\ref{j}), we can in principle calculate the charge
current at any time $t$ for any given initial instantaneous
eigenstate of the system. Noting Eqs.~(\ref{eq}) and (\ref{E}), we have
\begin{equation}\label{Jt}
 j(t)=B_1 \cos \left( \frac{e}{\hbar}Et \right) -B_2 \sin \left( \frac{e}{\hbar}Et
\right),
\end{equation}
where
\[ B_1=\frac{2t_{0}e}{\hbar}\sum_{m=1}^{N}\sin q_{m};\quad B_2= \frac{2t_{0}e}{\hbar}\sum_{m=1}^{N}\cos
q_{m}. \]

For the half-filling case, $N=L$, one can find that both $B_1$ and
$B_2$ are 0,  due to the complete cancellation in the sum,  and
hence $j(t)=0$.

However, away from half-filling, nonzero $B_1$ and $B_2$ will lead
to a nonzero $j(t)$, which oscillates with a phase frequency $\omega
=eE/\hbar$ and amplitude $\sqrt{B_{1}^{2}+B_{2}^{2}}$, the
so-called Bloch oscillation. The Bloch oscillation
now, similar to that in the noninteracting case, is an undamped one,
in spit of the infinity on-site interaction $U$.

The surviving Bloch oscillations under strongly correlated interactions is interesting. As we have mentioned, there have already existed
some discussions about the issue that electrons in strong correlated
systems could show dissipationless transport behaviors in the linear
response level \cite{In1,In2,In3,In4}. Our result is simple but obviously beyond the linear
 level, and can be viewed as a lively example when discussing the conductivity of Hubbard model.
 
An important question is that whether  such result is a special one  only in the 1D $U\rightarrow +\infty$ case.
 It is interesting to find more examples.

\subsection{Example for Bloch oscillations in finite $U$ cases}

We want to show that undamped Bloch
oscillations can still exist  in finite $U$ cases. An exact
discussion of this question in general sense seems impossible.
However, we find that  there do exist undamped Bloch oscillations for a
 finite $U$. A simple example is given via an analysis of the
current response of the special eigenstates constructed  by Yang and Zhang \cite{eta} for the  Hubbard model. The operator
 used,
$\zeta^{\dag}=\Sigma_{m=1}^{L}C^{\dag}_{m\uparrow}C_{m\downarrow}$,
which is found to be an eigen-operator of the Hamiltonian
(\ref{H1}), for any magnitude of $U$, satisfies
\begin{equation}\label{eta}
[\zeta^{\dag},H]=0.
\end{equation}

The operator $\zeta^{\dag}$, when acting on a state,  only transforms one of the
spin $\downarrow$ electrons in the system to a spin $\uparrow$ one.
It should be noted that if all the $N$ electrons in the system are
spin $\downarrow$ ones, the on-site $U$ will play no role. And then, the
time-evolution states are just the same as
that in the noninteracting case, and would show the same Bloch
oscillations. We can introduce a state of this kind and denote it by
$|\Psi_{0}(t)\rangle$. Then, due to Eq.~(\ref{eta}), the
state $|\Psi_{m}(t)\rangle=(\zeta^{\dag})^{m}|\Psi_{0}(t)\rangle$
($m\leq N$) is also a time-evolution state of Hamiltonian
(\ref{H1}), but belongs to a system with  $m$ spin $\uparrow$ electrons and
$N-m$ spin $\downarrow$ electrons. Since acting of
$\zeta^{\dag}$'s on $|\Psi_{0}(t)\rangle$ only changes the spin of
electrons from $\downarrow$ to $\uparrow$, the characteristic of
Bloch oscillations keeps. That is, time-evolution states of this
kind  do show undamped Bloch oscillations as that in the
noninteracting case,  completely disregarding the on-site $U$ in the
Hubbard model. Obviously, such result is also applicable to the two-
or three-dimensional (2D or 3D) case.

It should be noted that the number of  eigenstates   generated with the $\zeta^{\dag}$ operator is directly related to the particle number $N$ and the total site number $L$ in the system, and is in a scale of  $C_{L}^{N}$, which  generally is very large. Hence, the existence of undamped Bloch oscillations is extensive for Hubbard model and one should be cautious of their roles when discussing the conductance of the system. 

\subsection{More general cases}

At first glance, the undamped Bloch oscillations we obtained are directly related to the integrability of the model. It seems that such result is hard to be extended to more general cases, since it has been proposed that breaking of integrability would lead to damping of the charge current \cite{cc}. 
However, we want to show that the role the temporal periodicity of the model Hamiltonian playing in electric field is more important.
 
In an constant electric field $E$, when Peierls substitution \cite{PS} is adopted, the Hamiltonian and charge current operator for Hubbard model or other similar ones, such as $t-V$ and $t-V-W$ models \cite{cc},  are always periodic in time, 
 \begin{align*}
 & H(t+\tau)=H(t),\\
& \widehat{j}(t+\tau)=\widehat{j}(t)\equiv -c\frac{\partial H(t)}{\partial A}=\frac{1}{E}\frac{\partial H(t)}{\partial t},
 \end{align*}
where the period $\tau$ is directly related to the electric field, e.g., for 1D Hubbard model shown in Eq.~(\ref{H1}), $\tau=2\pi \hbar
 /(eE)$ . And $A(t)=-cEt$ as usual. 
 
 The evolution of states determined by such periodic Hamiltonian has been extensively studied (see, e.g.,  Refs.~\cite{WK1,WK2,WK3}).  Due to such periodicity,  the evolution operator $U(t,t_i)$ satisfies $U(t+\tau,t_i+\tau)=U(t,t_i)$, and we have 
 \begin{equation*}
 U(t,0)=U(t,n\tau)\cdots U(2\tau,\tau)U(\tau,0)=U(t^{-},0)U^n(\tau,0),
 \end{equation*}
 where $t^{-}\equiv t-n\tau \in [0,\tau ]$.
 
Further, we can introduce the eigenstates and eigenvalues of $U(\tau,0)$. 
Since $U(\tau,0)$ is a unitary operator, $U(\tau,0)^{\dagger}U(\tau,0)=\mathbf{1}$, the modulus of its eigenvalues must be 1. For each eigenstate $|m\rangle$, the corresponding  eigenvalue must be in the form of $\exp(i\phi_m)$. Namely, we have
\begin{equation*}
U(\tau,0)|m\rangle=\exp(i\phi_m)|m\rangle.
\end{equation*}

Obviously, these eigenstates as a whole form a complete set of states for the system.

The time evolution of each $|m\rangle$ is given by
\begin{equation}\label{mt}
|m(t)\rangle=U(t,0)|m\rangle.
\end{equation} 

Then we have 
\begin{equation*}
|m(t+\tau)\rangle=U(t+\tau,\tau)U(\tau,0)|m\rangle=\exp(i\phi_m)|m(t)\rangle.
\end{equation*} 

The instantaneous energy and charge current for these states satisfy
\begin{align*}
& E_{m}(t+\tau)=\langle m(t+\tau) |H(t+\tau)|m(t+\tau)\rangle=E_{m}(t),\\
& j_{m}(t+\tau)=\langle m(t+\tau) |\widehat{j}(t+\tau)|m(t+\tau)\rangle=j_{m}(t),
\end{align*}
namely, both are periodic in time. Then the charge current must be a constant or keep oscillating with a period $\tau$. However, noting that 
\begin{equation*}
 \frac{\partial E_{m}(t)}{\partial t}=\frac{\partial }{\partial t}\langle m(t) |H(t)|m(t)\rangle=Ej_{m}(t),
\end{equation*}
we find that, to guarantee the periodicity of $E_{m}(t)$, the current $j_{m}(t)$ can not be or approach a nonzero constant. 

 Additionally, the average of  $j_{m}(t)$ over a period gives
\begin{equation*}
\bar{j}_{m}=\frac{1}{\tau}\int^{t+\tau}_{t}j_{m}(t)=-\frac{1}{E\tau}\int^{t+\tau}_{t}\frac{\partial E_{m}(t)}{\partial t}=0,
\end{equation*}
namely, the oscillating of charge current must be an up-and-down one about zero.

Hence, for such complete set of states, the charge current, if nonzero, can only show an undamped oscillating behavior, which is very similar to Bloch
oscillations in the noninteracting case.

Such conclusion is a general one,  
and has nothing to do with the integrability or non-integrability of the model. 

As for the 1D $U\rightarrow +\infty$ Hubbard model, one can easily verify that the states shown in Eqs.~(\ref{pt}) and (\ref{mt}) are the same thing.

In the appendix, we consider a 2D model of electrons with $\delta$-function interactions in electric field, which can be viewed as a version of Hubbard
model in a continuous space. Such model is obviously non-integrable in the common sense. However, we find that the charge current can be exactly determined as $j(t) =j(0)+\frac{e^2}{m}ENt$,  in an undamped form disregarding the non-integrability of the model. Such undamped behavior of charge
current, exceeds our expectation once again, but can be viewed as a counterpart to the undamped Bloch oscillations of the Hubbard model, since both survive with the correlation interactions between electrons.

\section{Conclusion}
In conclusion, when an electric field appearing, the 1D Hubbard model in the $U\rightarrow
+\infty$ limit still can be exactly solved. And undamped 
Bloch oscillations are found to extensively exist not only in this case, but also in finite $U$ or higher-dimension cases, and hence can not be ignored on discussing the conductivity of Hubbard model.

The origin of such undamped-oscillating behavior of charge currents  is mainly due to  the temporal periodicity of the  Hamiltonian of a model in electric field, rather than to the integrability of the model itself.

 The undamped 
Bloch oscillations in Hubbard model, and the dissipationless charge current in the $\delta$-function-interaction model as we shown in the appendix, both can survive with the strongly correlated interactions between electrons, calling  for our more  caution about the effect of correlation interactions and the integrability of a model on the conductivity of the system.

Additionally, the unitary-transformation method we adopted can also be used to  study the time-evolution behavior of states in an electric field for other similar 1D models such as the strong-coupling $ t $-$ V $ model and SU($ N $) model of impenetrable fermions.

\begin{acknowledgments}
The author acknowledges financial support from Guizhou Provincial
Education Department (Grant No.~QJHKY[2016]314) and Guizhou
Provincial Science and Technology Department (Grant
No.~QKHLH[2014]7433).
\end{acknowledgments}

\appendix*

\section{Exact conductivity for a non-integrable model}

To seek more about 
the effect of integrability and 
correlation interaction of electrons on the charge current of the
system, we now discuss with a $\delta$-function interaction model, which is
not the  Hubbard model but can be viewed as a version of Hubbard
model in a continuous space. And we find that the behavior of  charge currents can be exactly discussed.

For non-integrability consideration,  We study the 2D case, namely, a system of electrons with $\delta$-function
interactions in a torus; and one will find that the extension to other dimensions is straightforward. The Hamiltonian is formally written as
\begin{multline*}
H=\sum_{k\sigma}\frac{P_k^2}{2m}
C^{\dag}_{k\sigma}C_{k\sigma} \\
+U\sum_{k_1,k_2,q}C^{\dag}_{k_1+q
	\downarrow}C^{\dag}_{k_2-q\uparrow}C_{k_2\uparrow}C_{k_1\downarrow}
\end{multline*}
where $k=(k_x,k_y)$ and $P_k=\left[ \hbar k_x+eA(t)/c,\hbar k_y\right] $, i.e., the electric field is along the $x$ direction, $E=-1/c A_t (t)$. The involved charge
current operator is
\[
\widehat{j}_x=\sum_{k\sigma}\frac{\hbar k_x}{2m}C^{\dag}_{k\sigma}C_{k\sigma}-\frac{e^2}{mc}A(t)\sum_{k\sigma}C^{\dag}_{k\sigma}C_{k\sigma}.
\]

Then the charge current for an evolving state $|\Psi(t)\rangle$ is
\[
j(t)=\langle \Psi(t)|\widehat{j}_x|\Psi(t)\rangle,
\]
and
\[
\frac{dj(t)}{dt}=\frac{1}{i\hbar}\langle
\Psi(t)|[\widehat{j}_x,H]|\Psi(t)\rangle+\langle \Psi(t)|\frac{\partial
	\widehat{j}_x}{\partial t}|\Psi(t)\rangle =\frac{e^2}{m}EN,
\]
where we have used the fact that $[\widehat{j}_x,H]=0$, and $N$ is the total
number of electrons in the system. Then,
\[
j(t) =j(0)+\frac{e^2}{m}ENt,
\]
where $j(0)$ is the current at $t=0$. Namely, a dissipationless charge
current appears with the electric field. The non-integrability of the model and interactions between electrons do not lead to any damping.  Our analysis can be easily
extended to other dimensions or cases with more
general particle-particle interactions.

\end{document}